\documentclass[aps, prl, twocolumn,  superscriptaddress, floatfix, showpacs]{revtex4}

\usepackage{graphicx}
\usepackage{dcolumn}
\usepackage{bm}
\usepackage{times}
\usepackage{color}
\usepackage[normalem]{ulem}

\begin{document}

\bibliographystyle{apsrev4-1}

\preprint{APS/123-QED}

\title{Light atom quantum oscillations in UC and US}

\author{Yuen Yiu}
\altaffiliation{author to whom correspondences should be addressed: E-mail:[yyiu@utk.edu]}
\affiliation{Department of Physics and Astronomy, University of Tennessee, Knoxville, TN 37996, USA}
\author{A.A. Aczel}
\altaffiliation{or [aczelaa@ornl.gov]}
\affiliation{Quantum Condensed Matter Division, Oak Ridge National Laboratory, Oak Ridge, TN 37831, USA}
\author{G.E. Granroth}
\affiliation{Neutron Data Analysis and Visualization Division, Oak Ridge National Laboratory, Oak Ridge, TN 37831, USA}
\author{D.L. Abernathy}
\affiliation{Quantum Condensed Matter Division, Oak Ridge National Laboratory, Oak Ridge, TN 37831, USA}
\author{M.B. Stone}
\affiliation{Quantum Condensed Matter Division, Oak Ridge National Laboratory, Oak Ridge, TN 37831, USA}
\author{W.J.L. Buyers}
\affiliation{Chalk River Laboratories, Canadian Neutron Beam Centre, National Research Council, Chalk River, Ontario, Canada, K0J 1J0}
\author{J.Y.Y. Lin}
\affiliation{ California Institute of Technology, Pasadena, California 91125, USA}
\author{G.D. Samolyuk}
\affiliation{Materials Science and Technology Division, Oak Ridge National Laboratory, Oak Ridge, TN 37831, USA}
\author{G.M.~Stocks}
\affiliation{Materials Science and Technology Division, Oak Ridge National Laboratory, Oak Ridge, TN 37831, USA}
\author{S.E. Nagler}
\affiliation{Quantum Condensed Matter Division, Oak Ridge National Laboratory, Oak Ridge, TN 37831, USA}

\date{\today}

\begin{abstract}

High energy vibrational scattering in the binary systems UC and US is measured using time-of-flight inelastic neutron scattering.   A clear set of well-defined peaks equally separated in energy is observed in UC, corresponding to harmonic oscillations of the light C atoms in a cage of heavy U atoms.  
The scattering is much weaker in US and only a few oscillator peaks are visible. We show how the difference between the materials can be understood by considering the neutron scattering lengths and masses of the lighter atoms.  
Monte Carlo ray tracing is used to simulate the scattering, with near quantitative agreement with the data in UC, and some differences with US.  
The possibility of observing anharmonicity and anisotropy in the potentials of the light atoms is investigated in UC. Overall the observed data is well accounted for by considering each light atom as a single atom isotropic quantum harmonic oscillator.

\end{abstract}

\pacs{63.20.dd, 63.20.Pw, 78.70.Nx}
\maketitle

\section{I. Introduction}

The complex electronic structure of uranium leads to a wide variety of unusual and diverse behavior in the uranium salts UX (X = C, N, P, S, As, Se, Sb, Te, Bi), and therefore the materials in this family have been the subject of many past investigations characterizing their magnetic and vibrational properties\cite{69_kuznietz, 74_wedgwood, 80_buyers, 81_buyers, 82_holden, 82_holden_2, 83_buyers, 86_jackman, 87_holden, 87_hughes}.  
The phonon spectra in the UX systems have been comprehensively studied, however there has been renewed  interest in more detailed calculations of their vibrational properties as certain members of the family, including UN and UC, are under active consideration for next generation nuclear fuels\cite{09_burkes, 11_yin, 13_mei}. 

Recently, new and unexpected features were discovered in the vibrational spectrum of UN\cite{12_aczel}.  The same single crystal of UN studied in earlier work\cite{77_dolling, 84_holden, 86_jackman} was re-examined via modern neutron time-of-flight (TOF) spectroscopy using the SEQUOIA\cite{06_granroth, 10_granroth} and ARCS\cite{12_abernathy} spectrometers at the Spallation Neutron Source, Oak Ridge National Laboratory. 
Specifically strong vibrational scattering was found at energies above the usual acoustic and optic phonon branches.
 A series of evenly-spaced, high energy modes was observed, and detailed quantitative analysis showed that these modes could be attributed to the nitrogen atoms behaving as independent, isotropic, 3D quantum harmonic oscillators (QHOs)\cite{12_aczel, 14_lin}. 
 Some of the features of the QHO modes, for example the intrinsic broadening, were consistent with predictions of a binary solid model\cite{83_lovesey}.  This model applies to systems with two different types of atoms with disparate masses and explains how the motion of the heavy atoms affects the QHO behavior of the lighter ones.

The observation of these well-defined modes at energies above the highest optic phonon in an ordered single crystal contrasts with the conventional view of vibrational response in crystalline solids.   Above the highest optic phonon modes the response is usually weak and relatively featureless\cite{dove_book}. Prior to the UN discovery, exceptions to this rule were generally  found in binary metallic hydrides\cite{87_ikeda, 92_kolesnikov, 94_kolesnikov, 91_elsasser, caputo_03}, for example in ZrH$_x$ systems, where hydrogen atoms occupy interstitial sites\cite{94_kolesnikov}. 
 However in those systems, the hydrogen modes usually exhibit significant anisotropic and anharmonic effects, mainly due to H-H interactions, crystalline anisotropy, and the diffusion of the H atoms\cite{87_ikeda, 92_kolesnikov, 94_kolesnikov, 91_elsasser}. 
 Typically only a few modes are observed.   
 Conversely, the nitrogen oscillations in UN show well-defined peaks up to the 10$^{th}$ order.

Known pre-requisite conditions for the QHO modes to be clearly observable in binary alloys include a large mass ratio between the light and heavy atoms, and weak interactions between light atoms. Although the aforementioned binary solid model provides some guidance, there is little experimental information on the dependence of this part of the vibrational response function on factors such as the mass ratio and atomic neutron cross-sections.  In this work, we investigate how these factors affect the high energy vibrational scattering in UX via time-of-flight neutron spectroscopy measurements on single crystals of UC and US. For UC, we find a series of well-defined high energy vibrational modes analogous to those observed for UN. Overall these modes are described well by a QHO model for the carbon atoms. Conversely, while the US data also shows evidence for the high energy vibrational modes, only the lowest few are observed and they are much weaker in intensity compared to those seen in UC and UN.   We discuss the reasons for these differences.  In addition we explore the effect of multiple scattering on the observed spectrum, and the possibility of extracting information on the directional dependence and anharmonicity of the light atom potential via measurements of the QHO modes.

\section{II. Inelastic neutron scattering}

 All neutron scattering measurements reported here were collected using the SEQUOIA\cite{06_granroth, 10_granroth} and ARCS\cite{12_abernathy} TOF Fermi chopper spectrometers at the Spallation Neutron Source of Oak Ridge National Laboratory. 
 The same depleted uranium single crystals of UC and US used in previous studies were investigated in this work\cite{85_duplessis, 71_smith}. Both samples had similar total volumes on the order of 1~cm$^3$. 
 For the neutron scattering experiments, each single crystal was mounted in an aluminum can and loaded in a closed cycle helium refrigerator. 
 All data were collected at $T$~$=$~4~K, with the $[HHL]$ scattering plane horizontal. 
 A Fermi chopper was used to obtain several different incident neutron energies, including $E_i$~$=$~80~meV, 250~meV, 500~meV, 700~meV and 800~meV. The details for each chopper setting are given in Table~\ref{table1}. 
 All datasets were normalized against a vanadium standard to account for variations of the detector response and the solid angle coverage. 
 Empty can measurements were performed at $T =$ 4 K and subtracted from the US datasets, however a small amount of Al that made up the sample mount was not perfectly accounted for. 
 We discuss below the implications of this imperfect background subtraction on the scattering observed in US. 
 A similar background subtraction was found to be unnecessary for the UC data because the intrinsic features of interest are intense. 
 
 
%

\begin{table}[htb]
\begin{center}
\caption{\label{table1}List of experimental conditions used to collect data for (a): UC, and (b): US, on SEQUOIA($E_i$ = 80 meV only) and ARCS($E_i$ = 80, 250, 500, 700 and 800 meV).  
Here $d_{FC}$ is the slit spacing of the Fermi chopper, $R_{FC}$  is the radius of curvature of the Fermi chopper, $\nu_{FC}$  is the frequency of the Fermi chopper, $\nu_{T0}$ is the frequency of the T0 chopper, and $\Delta E$ FWHM represents the corresponding instrumental energy resolution at zero energy transfer.}

(a) UC, $T$ = 4 K \\
\begin{tabular}{|c|c|c|c|c|}
\hline
$E_i$ (meV)       & 80   & 500   & 700  & 800         \\ \hline
$d_{FC}$ (mm)     & 1.5    & 0.5    & 0.5   & 0.5      \\
$R_{FC}$ (m)      & 0.58   & 1.53   & 1.53  & 1.53     \\
$\nu_{FC}$ (Hz)   & 120  & 480  & 600 & 600    \\
$\nu_{T0}$ (Hz)   & 90   & 180  & 180 & 180    \\   
$\Delta E$ FWHM (meV) & 9.0   & 18.2  & 26.1 & 31.4  \\ \hline
 
\end{tabular}


\medskip

(b) US, $T$ = 4 K \\
\begin{tabular}{|c|c|c|c|}
\hline
$E_i$ (meV)   & 80    & 250    & 500        \\ \hline
$d_{FC}$ (mm) & 3.6   & 0.5    & 0.5        \\
$R_{FC}$ (m)  & 1.53  & 1.53   & 1.53       \\
$\nu_{FC}$ (Hz) & 240 & 360    & 480        \\
$\nu_{T0}$ (Hz) & 90  & 180    & 180        \\    
$\Delta E$ FWHM (meV)  & 4.1   & 8.8  & 18.2  \\ \hline
\end{tabular}
\end{center}
\end{table}

\begin{figure}
\centering
\includegraphics[width=70mm]{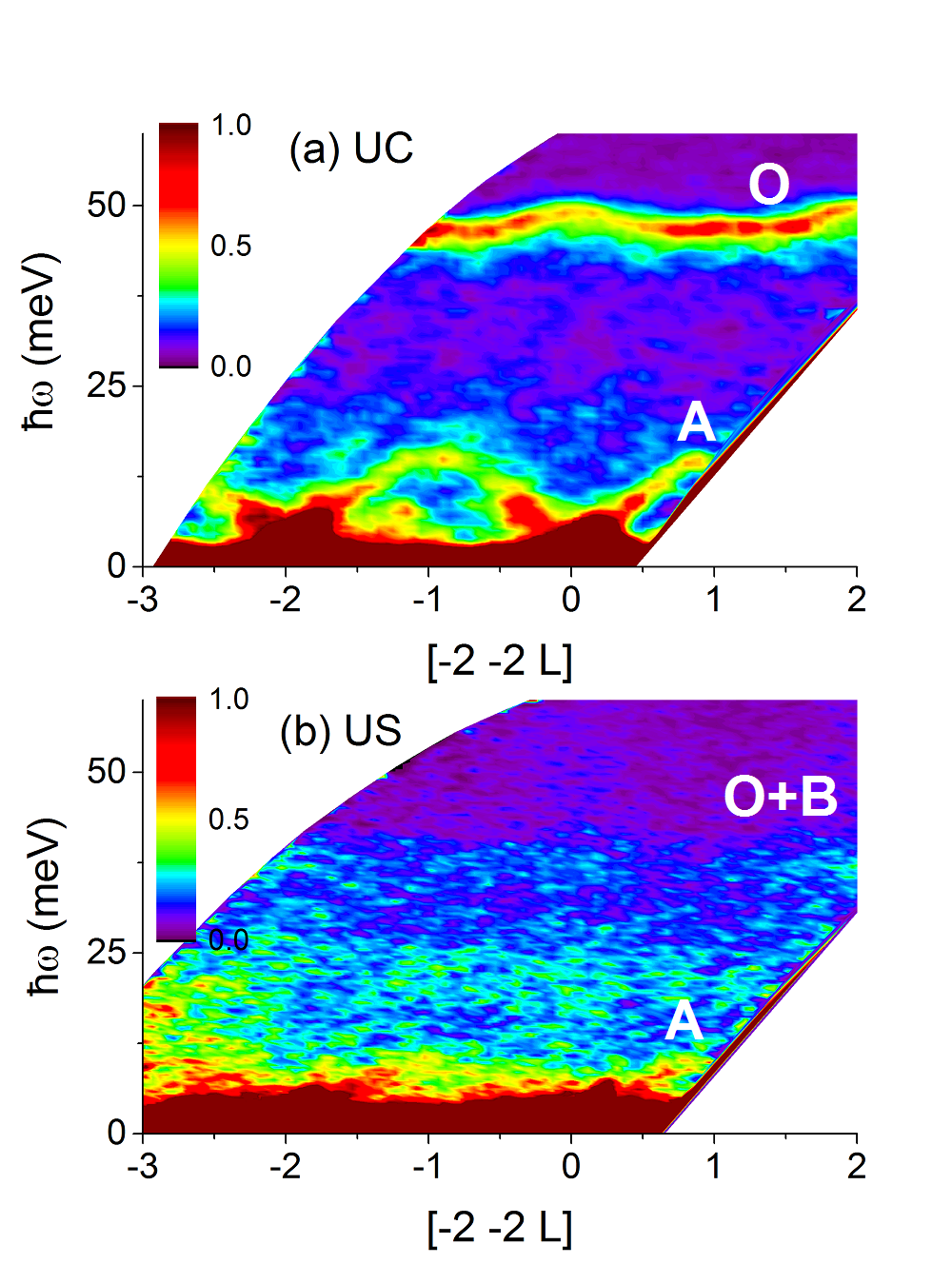}
\caption{\label{Fig1} The phonons along the [-2 -2 L] direction for (a) UC (ARCS) and (b) US (SEQUOIA) with $E_i~$~$=$~80~meV. In both cases, the acoustic and optic phonon modes are well-separated. The optic modes' signal to background ratio is much higher for UC as compared to US. The energies of the optic modes for UC and US are known to be 48 meV and 40 meV \cite{86_jackman}. The white text O and A indicate the general location for the optic and acoustic phonon modes, while B denotes significant background contribution.} 
\end{figure}

Figure~\ref{Fig1} shows representative phonon measurements with $E_i~$~$=$~80~meV for (a) UC from ARCS and (b) US from SEQUOIA along the [$\bar{2} \bar{2} L$] direction in reciprocal space. The results for the acoustic and optic phonons are consistent with those previously reported in the literature\cite{86_jackman, 71_smith}.  Both the acoustic and optic modes in UC are clearly visible with large signal compared to background.  On the other hand, the optic phonon modes of US in particular are only weakly visible in the data. The decreased intensity of these excitations can be understood by considering the one phonon structure factor, $g$, for the rocksalt structure. Along symmetry directions this can be written as \cite{74_wedgwood, 66_peckham}: 
\begin{equation} 
g^2=(\mathbf{Q\cdot\boldsymbol{\xi}})^2g'^2
 \end{equation}
with
\begin{equation} 
g'=\left(\frac{b_U d_U}{\sqrt{M}}\pm \frac{b_X d_X}{\sqrt{m}}\right)
 \end{equation}
where $\bf{Q}$ is the neutron momentum transfer, $\boldsymbol{\xi}$ is a unit vector describing the phonon polarization, $M$ is the mass of U, $m$ is the mass of the light atom X, $d_U$ and $d_X$ are the relative displacements of the U and the X ions, subject to the constraint $d_U^2+d_X^2 = 1$, and $b_U$ and $b_X$ are the neutron bound coherent scattering lengths\cite{74_wedgwood}. The sign between the two terms in Eq.~(2) depends on the ($HKL$) indices of the Brillouin zone, where (+) and (-) correspond to all even and all odd indices respectively.

For optical phonons, at the zone center $|d_U M|$~$=$~$|d_X m|$ and therefore $g'$ can be rewritten as:
\begin{equation} 
g'=\frac{mM}{\sqrt{m^2+M^2}}\left(\frac{b_U}{M^{3/2}}\pm \frac{b_X}{m^{3/2}}\right)
\end{equation}
In the limit of $M\rightarrow \infty$ :
\begin{equation} 
\lim_{M\rightarrow \infty }g'^2=\frac{b_X^2}{m}
\end{equation}

This expression indicates that the intensity of the optic modes is determined primarily by the mass and scattering length of the light atoms.  Under the assumption of  similar scattering lengths, the intensity of the optic modes decreases with increasing mass $m$ of the light atom. 
Table~\ref{table2} further emphasizes this point by comparing several parameters relevant to the phonon intensities of UC, UN, and US \cite{nist}. 
It is seen that the optic phonon cross section for US is more than an order of magnitude smaller than the comparable cross-sections for UC or UN.  Al is also included in the table since. It leads to significant background scattering in the experiments, as discussed below.

\begin{table}[htb]
\begin{center}

\caption{\label{table2}Relevant parameters for phonon intensities in UC, UN, US, and Al. \cite{nist}} 

\begin{tabular}{| c | c | c | c |}
\hline 
Atom & $m$ (amu) & $b (fm)$ & $b^2/m (fm^2/amu)$ \\ \hline
N & 14 & 9.4 & 6.31 \\
C & 12 & 6.6 & 3.63 \\  
S & 32 & 2.8 & 0.25 \\  \hline
U & 238 & 8.4 &  - \\ 
Al & 27 & 3.4 &   -  \\ 
\hline
\end{tabular}
\end{center}
\end{table}

\begin{figure*}
\centering
\includegraphics[width=\textwidth]{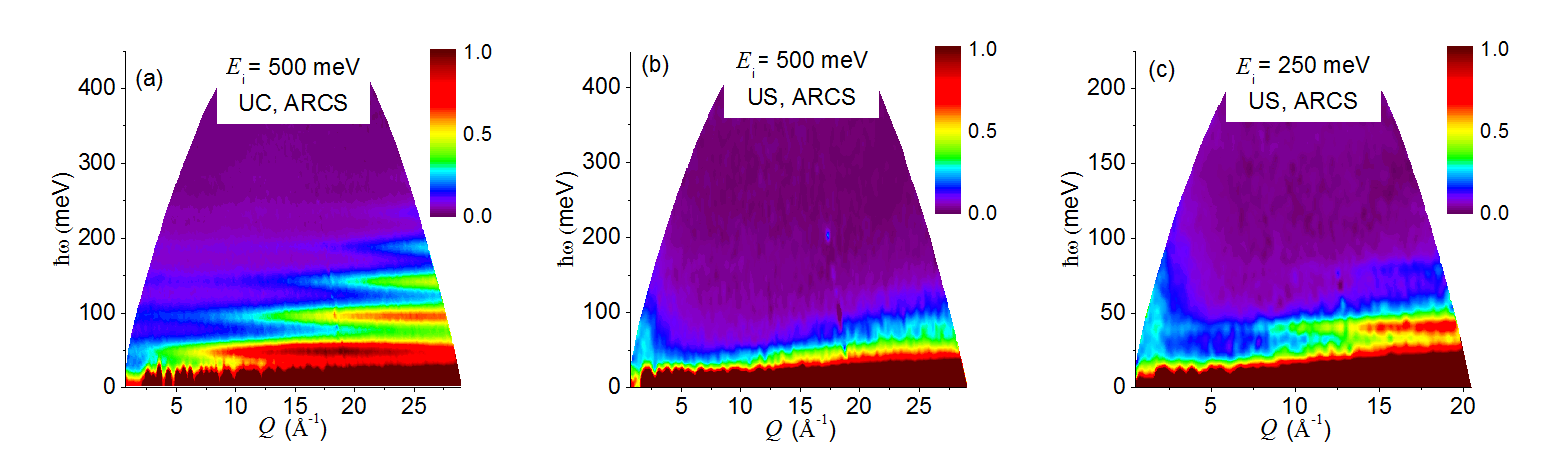}
\caption{\label{Fig2} Color contour plots of the experimental neutron scattering spectra for (a) UC, $E_i$~$=$~500~meV, (b) US, $E_i$~$=$~500~meV and (c) US, $E_i$~$=$~250~meV. The evenly-spaced, high energy vibrational modes are clearly visible in the UC plot, but much less pronounced  in the case of US. The small $Q$ inelastic scattering seen in US arises from magnetic excitations.}
\end{figure*}

Figure~\ref{Fig2} depicts the orientationally averaged, high energy response observed for both UC and US on ARCS. 
Several evenly-spaced vibrational modes are easily visible in the UC $E_i$~$=$~500~meV spectrum shown in Fig.~\ref{Fig2}(a).  In the US data with $E_i$~$=$~500~meV (Fig.~\ref{Fig2}(b)) the high energy vibrational modes are less obvious.  
On the other hand, some weak modes can be ascertained in the higher resolution $E_i$~$=$~250~meV data depicted in Fig.~\ref{Fig2}(c).  
In both materials the modes appear on inspection to be evenly spaced.

\begin{figure}
\centering
\includegraphics[width=73mm]{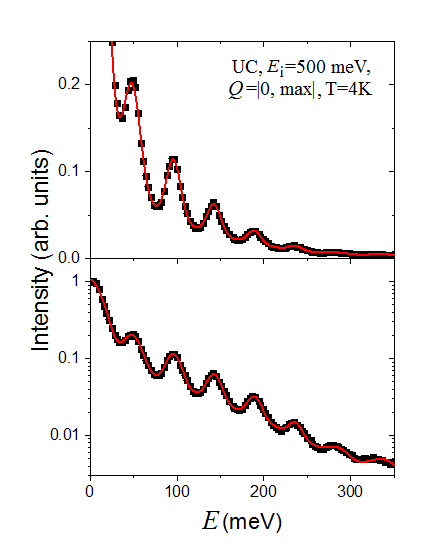}
\caption{\label{UC500} The $Q$-integrated intensity versus energy plot for UC at 4 K, with $E_i = 500$ meV. The upper plot shows the systematic decline of intensity for higher modes. The lower (log) plot shows that the QHO modes  remain visible for this dataset up to the 7th mode. The black points are from the experimental data and the solid red line is the fit to the data using Eq.~(5).}
\end{figure}

We first discuss the scattering seen in UC. Figure~\ref{UC500} shows the $E_i$~$=$~500~meV $Q$-integrated data for UC, 
with two different panels corresponding to linear (top) and log (bottom) y-axes.  Modes up to 7th order are clearly visible as peaks in the data.

To test whether these are evenly spaced the  $Q$-integrated dataset is fitted to the following functional form: 
\begin{equation}
I(E)=\sum_{n}T_{n}e^{-\frac{(E-E_{n})^2}{2\sigma _n ^2}}+B_{exp}e^{-\lambda _{B} E}+B_{0}.
\end{equation}
$T_n$ is a scale factor for each individual Gaussian peak, and $\sigma_n$ is the standard deviation for each peak. An empirical background consisting of a constant $B_0$ and decaying exponential $B_{exp}e^{-\lambda _{B} E}$ was incorporated into the fit. The decaying exponential term is expected from a simple diffusive model of multiphonon scattering \cite{nagler_91, perring_89}. $E_n$ are the mode positions that were fit independently and the last term incorporates all others sources of background.

The solid line in figure~\ref{UC500} represents the fitted curve from Eq.~(5). 
The fitted peak positions of the modes for all measured datasets are shown in Table~\ref{table3}, and are evenly spaced. 
The average spacing between the modes, incorporating all fitted datasets, was found to be $\hbar\omega_0$~$=$~48$\pm$1 meV for UC.  
The values are averaged by a weighted approach meaning $\hbar\omega_0 = (\hbar\omega_1 +  \hbar\omega_2/2 + \hbar\omega_3/3 + ...+ \hbar\omega_n/n)/n $. 
Values calculated from individual peaks, i.e.$\hbar\omega_n/n$ deviate from the average by less than 1 meV, which implies a highly harmonic potential.

\begin{table}[htb]
\begin{center}
\caption{\label{table3}Peak positions of the high energy vibrational modes (in meV) for UC from fits to the $Q$-integrated data. Please note that the error bars are statistical from the fitting of the data and do not account for instrumental resolution or systematic effects.}

(a) UC, $T$ = 4 K \\
\begin{tabular}{|c|c|c|c|}
\hline
$n\backslash E_i$ & 500meV   & 700meV   & 800meV                              \\ \hline
1      & 47.7(3)  &    -      &     -     \\
2      & 96.2(3)  & 98.1(2)  &    99.7(4)     \\
3      & 142.8(4) & 144.1(3) & 146.0(5) \\
4      & 189.1(6)   & 190.9(3) & 193.0(4) \\
5      & 235.6(1)   & 237.3(4) & 239.0(4) \\
6      & 283.8(36)   & 283.7(6) & 284.7(6)   \\ 
7      & -   & 329.9(7) & 331.0(1.0) \\
8      & -   & 372.9(1.2) & 375.5(1.7) \\ \hline
\end{tabular}
\\ average $\hbar\omega_0$: 48$\pm$1~meV
\end{center}
\end{table}

The fact that the spectrum is characterized by a series of evenly-spaced modes suggests that the light atoms behave like independent 3D QHOs similar to UN\cite{12_aczel}. 
This hypothesis can be tested quantitatively by comparing the relative intensities of the modes to the known dynamical structure factor for the $n^{th}$ mode of a QHO at low $T$, which has a simple analytical form given by: 
\begin{equation}
S_{n}(Q,\omega )=\frac{D}{n!}\left(\frac{\hbar Q^2}{2m\omega _{0}}\right)^n exp\left(\frac{-\hbar Q^2}{2m\omega _{0}}\right)\mathcal{F}(\omega)
\label{S_nqw_equ}
\end{equation} 
where $\mathcal{F}(\omega)$~$=$~$\delta(\hbar\omega -n\hbar \omega _{0})$ and $D$ is a constant.

Assuming that one has a carbon atom QHO with the measured energy spacing of $\hbar \omega_0$~$=$~48~meV and a mass of 12.01 a.m.u. the calculated value of $\hbar/2m\omega _0$ is 0.0036~\AA$^2$.  
Given the range of $Q$ measured and the $Q$ dependence in equation (6), it is surprising that the higher order modes are visible in figure 3. This can be understood by considering that  $S_{n}(Q, \omega)$ only accounts for single scattering events.  
With equally spaced modes, however,  the contributions from multiple scattering will also peak at energies corresponding to mode positions, and as discussed below the multiple scattering contributes to intensity at low $Q$. 

\begin{figure}
\centering
\includegraphics[width=82mm]{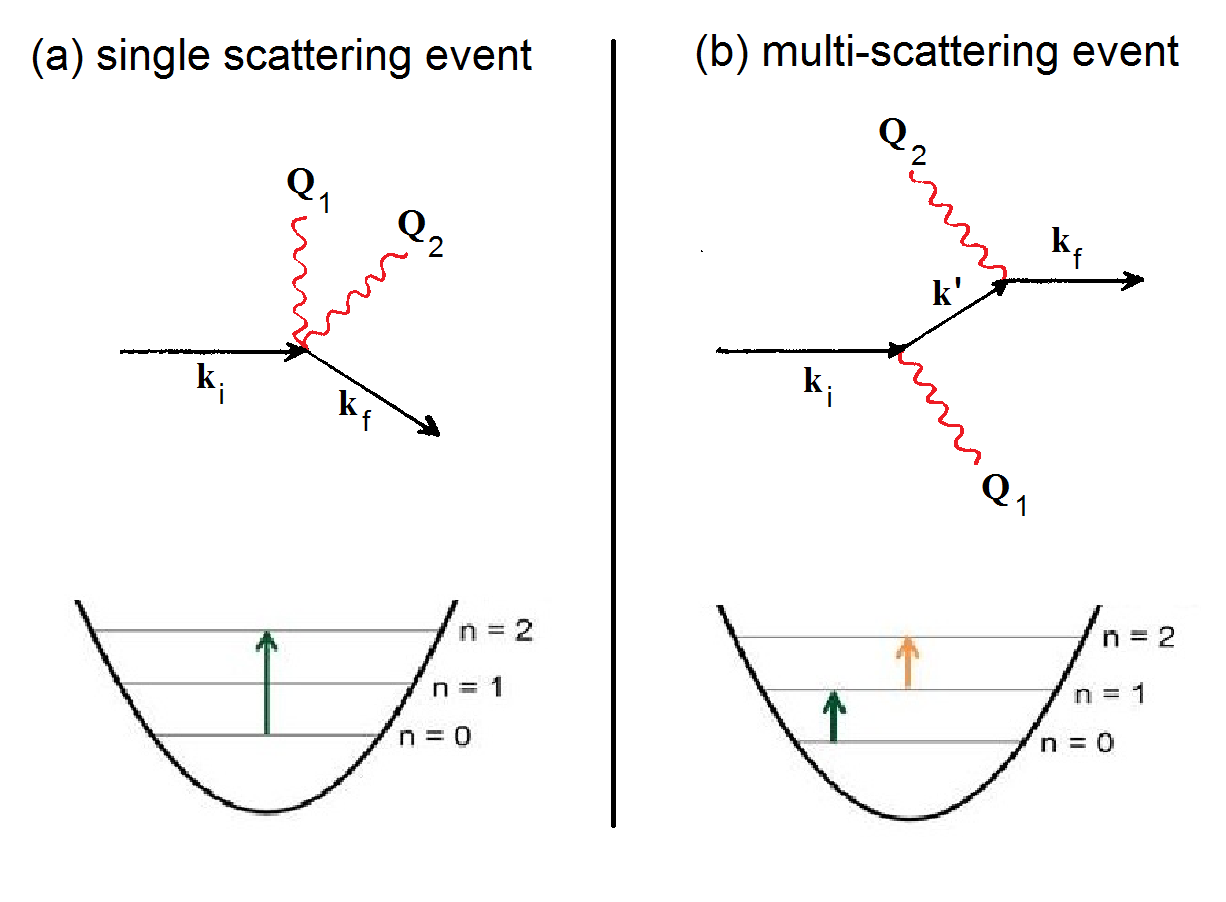}
\caption{\label{Fig3} Schematics showing the difference between (a) single scattering process and (b) multiple scattering processes leading to the observed neutron scattering spectra for UC, UN, and US. $k_{i}$ and $k_{f}$ refer to the incident and final neutron momenta.  
In (b) values of $\bf{Q_1}$ and $\bf{Q_2}$ were chosen such that the total momentum transfer $\bf{Q}$~$=$~0 but the scattering intensity is non-zero (see text for details).}
\end{figure}

We consider the meaning of a single scattering event in both QHO language and an alternate, but equivalent description based on creating Einstein phonons of fixed frequency.  
At $T=0$ an inelastic neutron scattering event, at an energy of $n\hbar\omega _0$ in QHO language, represents a transition from the QHO ground state to the $n^{th}$ eigenstate.  
In Einstein mode language it consists of a single scattering event, with total wavevector transfer magnitude $Q$, that creates $n$ Einstein phonons, each of energy $\hbar\omega _0$.  
In either case the scattering cross-section contains a factor $Q^{2n}$. 
On the other hand, a multiple scattering event in QHO language corresponds to a set of multiple transitions from the ground state to the $n^{th}$ excited state, while in the phonon description several different scatterings occur creating in aggregate $n$ Einstein phonons.  The $j^{th}$ scattering event contributes a factor $Q_{j}^2$ to the total observed cross-section.  
These processes are depicted in Figure~\ref{Fig3} for the case $n=2$.  The single scattering event shown in Figure~\ref{Fig3} (a) will have an intensity $I \sim Q^4$ and a requirement that the vector sum of the individual Einstein phonon momenta $\bf{Q_1}$ and $\bf{Q_2}$ has a magnitude equal to $Q$.  
On the other hand, in the multiple scattering event shown in Figure~\ref{Fig3} (b) each created Einstein phonon contributes a separate factor to the measured intensity which is therefore proportional to $(Q_1Q_2)^2$. 
The total net vector momentum transfer $\bf{Q}$ measured in the experiment can take on a wide range of values, including $\bf{Q}=0$ as depicted in the figure. 
The average over all possible multiple scattering combinations can lead to a cross-section that is independent of $Q$. Therefore the observed intensity of the higher order peaks at small $Q$ arises principally from multiple scattering.

With this in mind, we fit the constant-$E$ cuts of the UC data to a modified expression of $S(Q, \omega)$ for QHOs \cite{12_aczel}, given by:
\begin{equation}
S_{n}(Q,\omega )=A_n Q^{2n} exp(-CQ^2)+B_n
\end{equation}
Here $C$ is $\hbar/2m\omega_0$ and a $Q$-independent $B_n$ term is included to account for the multiple scattering described above. 
For the ideal QHO model, $A_n/D$~$=$~$C^n/n!$.  By relaxing this constraint and allowing $A_n$ and $C$ to be independent parameters in the data fitting, one can gain a sense of how much the UC data deviates from the ideal QHO limit. 

\begin{figure}
\centering
\includegraphics[width=70mm]{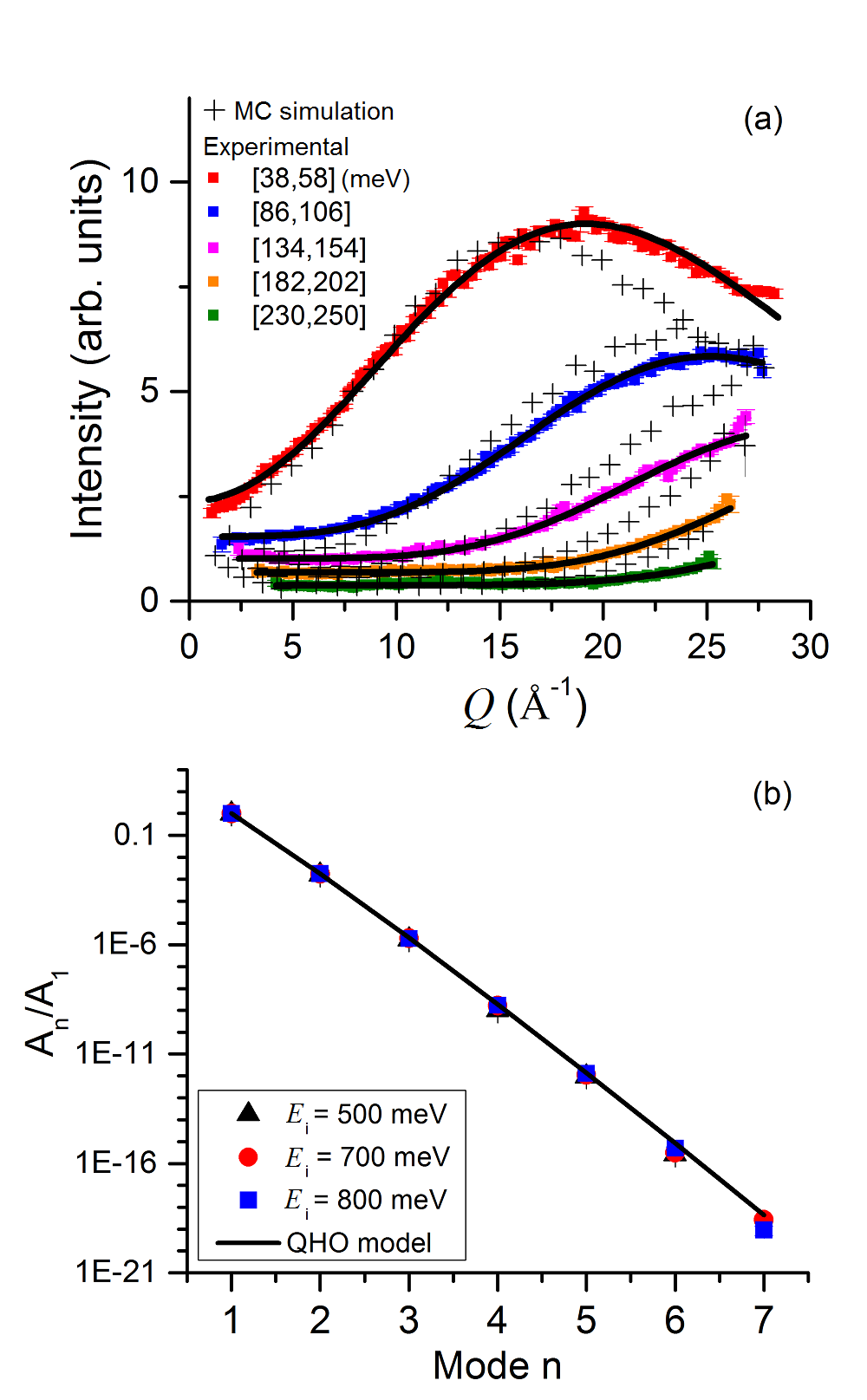}
\caption{\label{Fig4} (a) The $Q$-dependence of the intensity for the $n$~$=$~1-5 UC oscillator modes with $E_i$~$=$~500~meV, integrated over the energy ranges shown in the legend.  The solid lines are fits to the QHO model described in the text, while the $(+)$ symbols represent constant-$E$ cuts from the MC simulations presented in section III. (b) The $A_n$ coefficients for the $n$~$=$~1-5 modes normalized by the values of $A_1$. The coefficients are extracted from fits to the $E_i$~$=$~500~meV (black triangles), 700~meV (red circles) and 800~meV (blue squares) ARCS datasets. The solid line is the prediction of the QHO model with $\hbar \omega_0$~$=$~48~meV and $m$ corresponding to the mass of a carbon atom. The ratio is plotted on a logarithmic scale and spans almost 20 orders of magnitude.}
\end{figure}

Figure~\ref{Fig4}(a) depicts constant-$E$ cuts from the 500~meV dataset for the $n$~$=$~1-5 modes, centered about $n\hbar \omega_0$ for the $n^{th}$ mode (20~meV integration range). Similar cuts were made for the $E_i$~$=$~700~meV and 800~meV datasets (not shown). Figure~\ref{Fig4}(b) plots the ratios of the fitted parameters $A_n/A_1$ for the $E_i$~$=$~500~meV, 700~meV, and 800~meV datasets. The solid line indicates the prediction for the QHO model with $\hbar \omega_0$~$=$~48~meV, with the non-integer $n$ values interpolated by using $\Gamma (n+1)$ to calculate $n!$. 
As in the case of UN, there is excellent agreement between the QHO model and the data over 20 orders of magnitude. 
The experimental $C$ value, corresponding to the zero point motion of the oscillator, also agrees well with the model. 
By incorporating all three datasets into the fit, the average value of $C$ was found to be 0.0031(1)~\AA$^2$, which is close to the calculated value for the ideal QHO of 0.0036~\AA$^2$ discussed above. These findings provide strong confirmation that the high energy vibrational modes observed in UC correspond to quantum oscillations of the carbon atoms in the system.

\begin{figure}
\centering
\includegraphics[width=73mm]{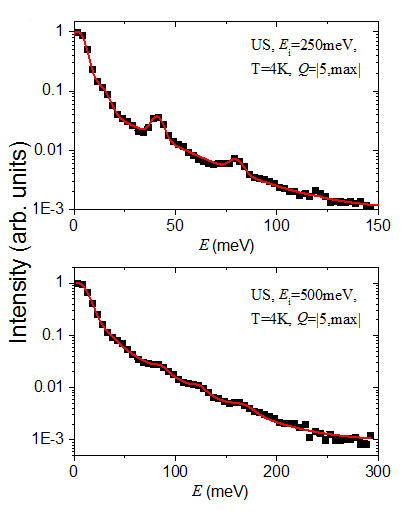}
\caption{\label{US500} The $Q$-integrated intensity versus energy plot for US at 4 K, with $E_i = 250$ meV and $E_i = 500$ meV. The QHO modes are much less intense compared to those observed in UC, with only 2 modes observable for $E_i=250$ meV and 4 for $E_i = 500$ meV. The black points are from the experimental data and the solid red lines are the fits to the data using Eq.~(5).  Please note that the $Q$-integrated range for the US data is $Q>5$ \AA$^{-1}$ only. This is done to avoid contributions from the magnetic excitations at low $Q$ as seen in Fig. 2 (b) and (c).}
\end{figure}

Figure~\ref{US500} shows the $Q$-integrated data for US with $E_i$~$=$~250~meV and $E_i$~$=$~500~meV, including fitted curves using Eq.~(5). 
The same fitting approach, described earlier for UC, was also used for the US data. However due to the weaker mode intensities in US, it is not possible to fit as many peaks in the US dataset.  Nonetheless the modes are evenly spaced, similar to UC, with an averaged $\hbar\omega_0$ of 41(1)~meV. The fitted peak position of each individual mode is shown in Table~\ref{tableUS}.  

\begin{table}[htb]
\begin{center}
\caption{\label{tableUS}Peak positions of the high energy vibrational modes (in meV) for US from fits to the $Q$-integrated data. Please note that the error bars listed are from the fitting of the data and do not account for instrumental resolution.}

US, $T$ = 4 K \\
\begin{tabular}{|c|c|c|c|}
\hline
$n\backslash E_i$ & 250meV   & 500meV            \\ \hline
1      & 40.9(3)    & -     \\
2      & 79.9(9)     & 85.6(2.4)    \\
3      & -    & 122.9(3.6)      \\
4      & -    & 166.7(5.8)     \\
 \hline
\end{tabular}
\\ average $\hbar\omega_0$: 41$\pm$1~meV
\end{center}
\end{table}

The weak intensities of higher QHO modes in US can be largely explained by the phonon structure factor of the rocksalt structure, as discussed earlier by  Eq.~(3) and (4), and Table~\ref{table2}. In short the greater mass and shorter scattering length of sulfur results in a much smaller $b^{2}/m$ compared to carbon or nitrogen.  At small $Q$ the scattering at the $n^{th}$ mode position is dominated by multiple scattering processes sequentially creating single optic phonons. The observed cross-section is therefore proportional to $(b^{2}/m)^n$.  For this reason the higher QHO modes appear much weaker in US compared to UC.

The correction for the Al background scattering is also a significant complication for the observation of QHO modes in US.  The QHO cross-section for each mode reaches a maximum at the value of $Q$ where the mode energy corresponds to the energy of the recoil scattering, which is given for an atom of mass $m$ by $E_{recoil}=\hbar^2Q^2/2m$.  As seen in Table II the masses of Al and S are relatively close, so the recoil scattering from the Al is strong near the position where one expects to see the scattering due to the QHO modes of S.  Altogether, quantitative analysis of the intensity of the QHO modes in US is more difficult.

\section{III. Monte Carlo ray tracing simulations}

Monte Carlo (MC) ray tracing simulations of both the UC and US ARCS neutron scattering experiments were performed to gain a better understanding of the various factors contributing to the observed scattering intensity. 
This approach worked exceptionally well to describe the neutron scattering spectra observed for UN\cite{14_lin}. 
The ability to model more realistic instrument and sample configurations is ever increasing\cite{14_lin_2, Farhi_14}.
The simulations used the MC ray tracing program Monte Carlo Virtual Neutron Experiment (MCViNE), which was based off Mcstas\cite{Lefmann_99,Willendrup_04} and developed in the Distributed Data Analysis for Neutron Scattering Experiments (DANSE)\cite{fultz_website} software development project. 
In each simulation, the sample was modeled as a cube with a 1~cm$^3$ volume, approximating the shape of the sample used in the measurement. 
The configuration files for the simulations were created through a series of simple modifications to the files used for the UN MC simulations described in detail in Ref.~\cite{14_lin}, allowing for a straightforward extension of those calculations to UC and US.   

Intrinsic broadening of the QHO modes is included in the simulation, as it is observed experimentally and predicted by the binary solid model\cite{83_lovesey}. The broadening arises from the fact that the heavy U atoms are not completely stationary, and can be modeled by replacing the Dirac-Delta function in the QHO expression with a Gaussian: 
\begin{equation}
\mathcal{F}(\omega)=exp\left(-\frac{(\hbar \omega-n \hbar \omega_o)^2}{2\Gamma^2(T)}\right)
\end{equation}
where the Gaussian width $\Gamma(T)$ is a function of temperature and given by:
\begin{equation}
\Gamma^2(T) = \frac{\hbar^2Q^2}{2M} \int_0^\theta du Z(u) u~coth\left(\frac{\hbar u}{2 k_B T}\right)
\end{equation}
M here is the mass of the heavy atom, in this case the U atom. $Z(u)$ is the acoustic phonon density of states calculated with a Born-Van Karmen model and $\theta$ is the maximum band frequency for the acoustic phonon modes.

\begin{figure*}
\centering
\includegraphics[width=\textwidth]{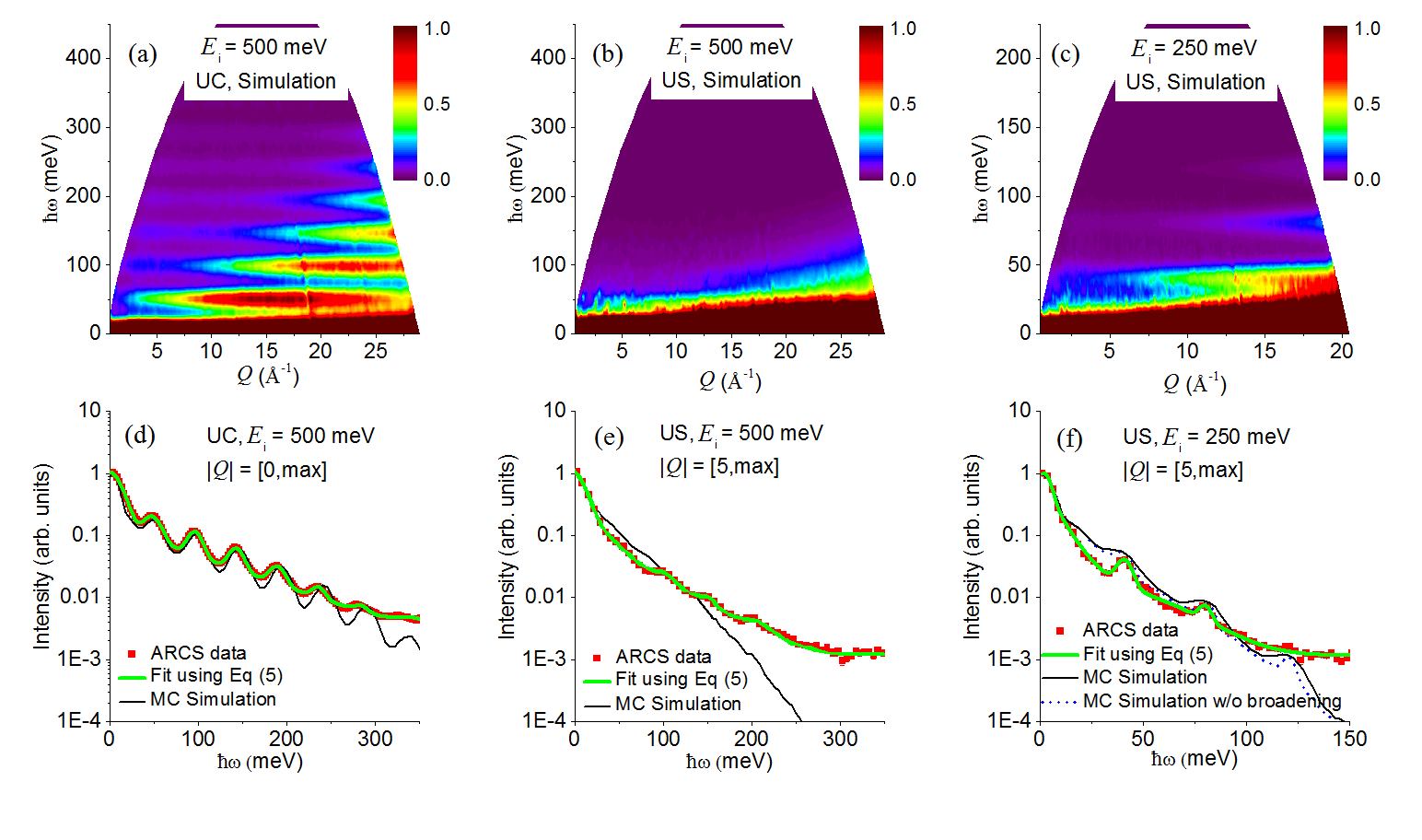}
\caption{\label{6panels}Color contour plots from the Monte Carlo ray tracing simulations described in text are shown here for (a) UC, $E_i$~$=$~500~meV, (b) US, $E_i$~$=$~500~meV and (c) US, $E_i$~$=$~250~meV.  The simulations that incorporate intrinsic broadening of the QHO modes agree well with the data for UC, but not for the case of US. (d), (e), and (f) show the corresponding $Q$-integrated plots for the three experimental data sets, and compare them directly to the results from the simulations, with solid green lines showing the fitted curve for each of the experimental data using Eq.~(5). (f) also shows MC simulation results both including and excluding intrinsic broadening of the QHO modes.}
\end{figure*}

A Monte Carlo ray tracing simulation for UC with $E_i$~$=$~500~meV is shown as a color contour plot in Fig.~\ref{6panels}(a) and the $Q$-integrated result is presented in Fig.~\ref{6panels}(d). This simulation is analogous to the one reported in Ref.~\cite{14_lin} for UN and includes various sample kernels accounting for elastic scattering, QHO scattering (from carbon atoms), acoustic phonon scattering, and all forms of multiple scattering arising from the processes described above. Figure~\ref{Fig4}(a) described earlier includes cuts from the $E_i$~$=$~500~meV MC simulation as (+) symbols with the experimental data, where $C$ is fixed at the value of 0.0036~\AA$^2$ corresponding to a carbon QHO with $\hbar \omega_0$~$=$~48~meV and a mass of 12.01 a.m.u.


Monte Carlo simulations for US, incorporating sample kernels with the same ingredients as for UC, are shown as color contour plots in Fig.~\ref{6panels}(b) and (c) with intrinsic broadening included. 
The $Q$-integrated results are also presented in Fig.~\ref{6panels}(e) and (f). For (f), data simulated without intrinsic broadening is also included.  At low signal levels the measured data approaches a constant value, above the simulations. This indicates that there is a background contribution, limiting the measurement sensitivity, that is not captured in the Monte Carlo simulations, nor by the Al can subtraction. The additional background could arise from a combination of effects including imperfect Al background subtraction, multiple scattering involving the magnetic response or events partially external to the sample, and possibly a contribution from fission neutrons in the sample that moderate in the instrument shielding.


When comparing the high energy vibrational modes of UC, UN, and US it is apparent that the differences in the U:X mass ratio (U:C~$=$~19.8, U:N~$=$~17, U:S~$=$~7.4) lead to more obvious QHO behavior in the lighter atoms.  As the U:X mass ratio decreases, the dispersion of the optic phonon modes has been shown to increase systematically\cite{74_wedgwood}, shifting the vibrational behavior of the light atoms further away from the localized QHO picture. A smaller U:X mass ratio has also the effect of pushing the optic phonon frequencies down closer in energy to the acoustic modes, so any resulting QHO excitations tend to be more closely spaced and therefore harder to resolve for a given instrumental energy resolution.

\section{IV. Anharmonicity and anisotropy}

\begin{figure*}
\centering
\includegraphics[width=\textwidth]{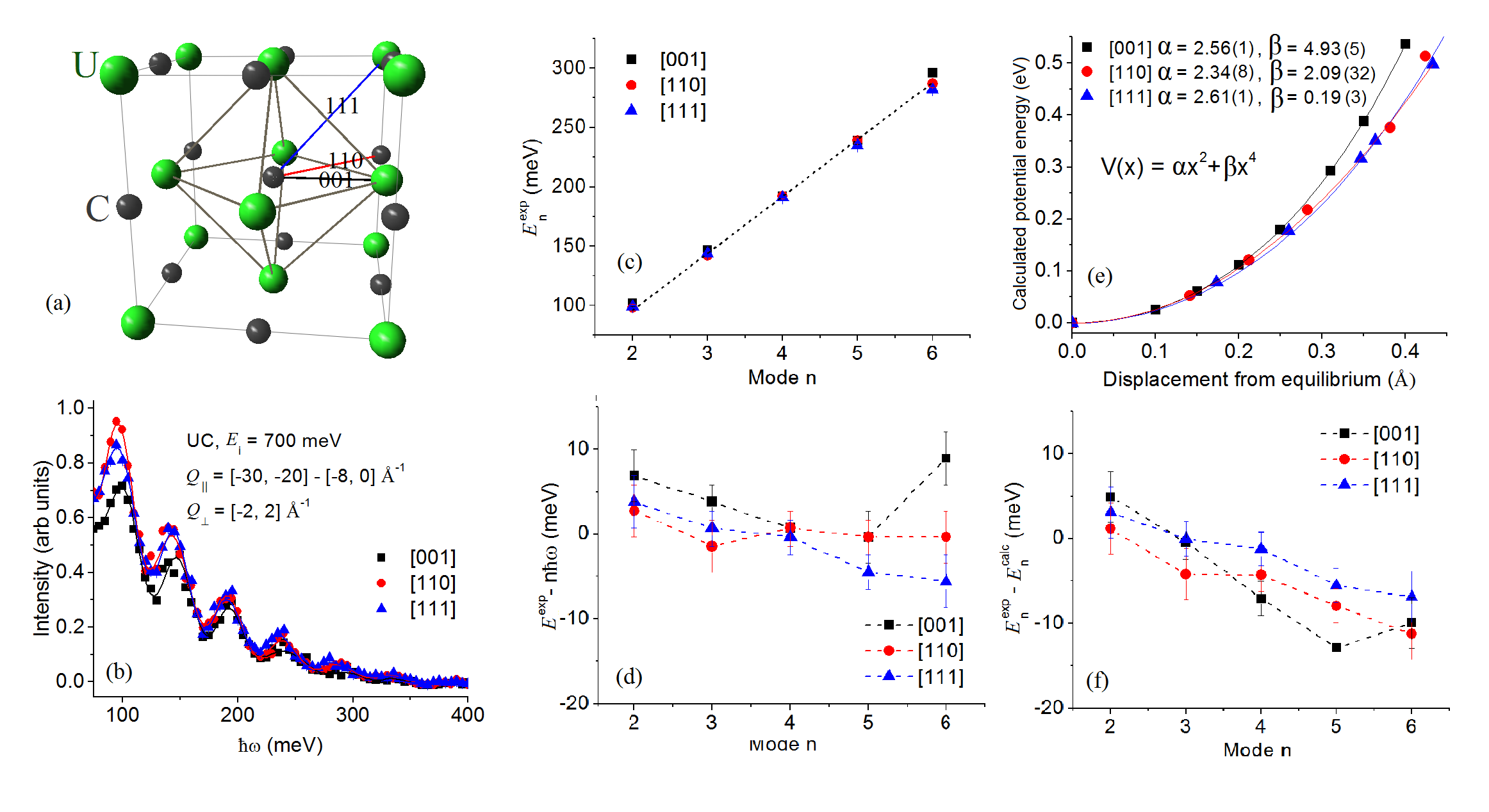}
\caption{\label{Fig5} (a) The crystal structure of the binary uranium systems UX. The light X atom is in an octahedral cage of U atoms.  The colored lines indicate the three primary crystallographic directions, [001], [110] and [111]. The carbon atom nearest neighbor distances in these directions are in the ratio of 1:$\sqrt{2}$:$\sqrt{3}$. (b) Constant-$E$ difference cuts along the major crystallographic directions [001], [110], and [111] for UC. The solid curves are fits of the data to Eq.~(5). (c) Peak positions along the three different directions taken from the fits in (b). The dotted line corresponds to the expected peak positions for an isotropic QHO. (d) The difference between the peak positions of the modes from our UC data and a perfectly harmonic potential. (e) Local potential of the carbon atoms calculated by DFT. The solid curves are fits to the anharmonic function V(x) given in the text. (f) The difference between the peak positions of the UC data and the result from the DFT calculations.}
\end{figure*}

All QHO data presented to this point for both UC and UN\cite{12_aczel} have been powder-averaged, so only $Q$ is considered and not $\bf{Q}$. The reason this approximation works so well is because the modes are highly isotropic. However, in principle the spacings of the modes could reflect both anisotropy and anharmonicity in the local potential.  To see whether these effects are experimentally observable in the QHO modes of UC, we collected a $E_i$~$=$~700~meV dataset with very high statistics along three crystallographic directions, [100], [110], and [111]. The C atoms have nearest neighbor distances along these directions in the ratio of 1:$\sqrt{2}$:$\sqrt{3}$, as one can see from considering the crystal structure shown in Fig.~\ref{Fig5}(a). The effects of anisotropy would be manifested by different mode spacings for the different directions, while anharmonicity would be apparent by different mode spacings as a function of energy.  Unfortunately, both of these features will be masked by the effect of multiple scattering, since at any value of $Q$ it tends to cause peaks to appear at locations corresponding to $n\hbar\omega_0$.  As a result, a straightforward plotting of oscillator peaks along different directions does not show any significant anisotropy or anharmonic effects, and additional analysis is needed to account for the multiple scattering.

As an effort to minimize the effect of multiple scattering present in the single crystal data, we take advantage of the fact discussed earlier that the small $Q$ scattering for the higher modes is totally dominated by multiple scattering, and use the low $Q$ data as a pseudo-background, as described here:  For each of the three directions we integrate over a range perpendicular to that direction given by $|Q_\perp|$ = [-2,2] \AA$^{-1}$.  As a signal we use the high $Q$ part of the scattering defined by 20~$<$~$Q_\parallel$~$<$~30~\AA$^{-1}$, and as a background we utilize the region 0~$<$~$Q_\parallel$~$<$~8~\AA$^{-1}$.  The subtraction of these two leads to constant-$Q$ difference cuts along the three directions as shown in Fig.~\ref{Fig5}(b). Note that the $n=1$ peak is not sufficiently resolved in the $E_i$~$=$~700~meV data to be included here.

\begin{table}[htb]
\begin{center}
\caption{\label{table4}Peak positions of the QHO modes (in meV) extracted from fitting difference cuts of the $E_i$~$=$~700~meV single crystal UC data along three major crystallographic directions. The form of the difference cuts is explained in the text.}
\begin{tabular}{|c|c|c|c|}
\hline
n & [001]    & [110]    & [111]    \\ \hline
1 & -       & -      &   -     \\
2 & 102(3) & 98(3)  & 99(3)  \\
3 & 147(2) & 142(3) & 144(3) \\
4 & 192(2) & 192(2) & 191(2) \\
5 & 239(3) & 239(2) & 235(2) \\
6 & 296(3) & 287(2) & 282(3) \\ \hline

\end{tabular}
\end{center}
\end{table}


The difference cuts are fit to Eq.~(5) to determine mode positions along the [001], [110], and [111] directions; fits are indicated in Fig.~\ref{Fig5}(b) by the solid curves, and the fitted peak positions are listed in Table~\ref{table4}.  These are also shown in Fig.~\ref{Fig5}(c), with the mode positions determined from the data ($E^{exp}_n$) and the dashed curve representing the expectation for an ideal carbon QHO with $\hbar\omega_0$~$=$ 48 $\pm$1 meV. Fig.~\ref{Fig5}(d) shows the difference between the mode positions for the ideal model (n$\hbar \omega_0$) and the experiment. The error bars are statistical and are larger than those for fits to the powder averaged data, presumably due to the lower number of counts in this single crystal dataset. From the fits, the average oscillator energies are given for the different directions as $E_{001}$=49.0(6)meV, $E_{110}$=48.0(6)meV, and $E_{111}$=47.9(6)meV.  There is no compelling evidence for anisotropy in the potential although the experimental data along the [001] direction did exhibit a slightly higher frequency than those along the other directions. This observation is compatible with density functional theory (DFT) calculations discussed below.  

To gain insight into any expected anisotropy and anharmonicity of the oscillator modes in UC, we have used DFT to calculate the potentials seen by the C atoms (see the appendix for details).  The calculated potentials for the three principal directions are shown by the symbols in Fig.~\ref{Fig5}(e). The potential in each direction is described well by the following expression:
\begin{equation}
V(x) = \alpha x^2 + \beta x^4
\end{equation}
with the results shown by the solid lines in the figure. The values determined for $\alpha$ and $\beta$ are also indicated.  The calculation shows a small amount of anisotropy, and also indicates that the anharmonicity should be maximized along the [001] direction. The latter result suggests that the harmonic approximation becomes less valid for in the direction with the shortest nearest neighbor distance. Indeed the fitted $\beta$ for [001]  is more than twice the magnitude of the $\beta$s along the other directions..

Assuming that the $x^4$ term is small, the energy levels corresponding to this potential can be calculated according to first order perturbation theory \cite{griffiths_book}, with the $n^{th}$ energy level corresponding to:
\begin{equation}
E_n =\hbar \omega_0 (n + 1/2) + \left ( \frac{\hbar}{2m\omega_0}\right )^2 \beta(6n^2+6n+3)
\end{equation}
and the energy difference between the $n^{th}$ level and the $n$~$=$~0 ground state given by: 
\begin{equation}
E_n - E_0 = n\hbar\omega_0 + 6\beta C^2(n^2+n)
\end{equation}
with $C$~$=$~$\hbar/2m\omega_0$. 

Assuming that the first mode occurs at 48 meV, the energies of higher QHO levels along each of the 3 directions can be calculated using Eq.~(12) along with each direction's corresponding $\alpha$ and $\beta$ shown in Fig.~\ref{Fig5}(e).  Fig.~\ref{Fig5}(f) displays the difference between the oscillator mode positions determined from the experiment and the value $E^{calc}_n$ from first order perturbation theory.  The downward slope indicates that the experimentally determined anharmonicity is not as large as the estimate.  One must bear in mind that despite the attempt to correct for multiple scattering in the data, it is greater at higher order peaks and therefore could still mask what is a rather subtle effect.

\section{V. Conclusions}

For the binary crystal UC, time-of-flight neutron scattering measurements reveal a series of well-defined, equally spaced, high energy vibrational modes that can be attributed to quantum harmonic oscillator behavior of the carbon atoms in this system. Measurements of the QHO modes along the high-symmetry [001], [110], and [111] directions reveal that these excitations are characterized by at most only a very small amount of anharmonicity and anisotropy. Similar time-of-flight neutron scattering data for US also shows evidence for the high energy vibrational modes, but only a few are clearly observed and they are much weaker in intensity. The difference can be understood by considering the U:X mass ratio, the quantity $b^2/m$ that characterizes the scattering strength of the modes, and the interference of Al and S recoil scattering with the signal in US.  Some progress has been made in modeling the various contributions to the scattering, but a better way to handle multiple scattering will be needed to accurately measure anharmonicity and anisotropy in the QHO modes.

\section{Acknowledgments}
This research was supported by the US Department of Energy, Office of Basic Energy Sciences.  A.A.A., G.E.G., D.L.A., M.B.S. and S.E.N. were fully supported and J.Y.Y.L was partially supported by the Scientific User Facilities Division. G.D.S. and G.M.S. were fully supported by the Materials Sciences and Engineering Division. Neutron scattering experiments were performed at the Spallation Neutron Source, which is sponsored by the Scientific User Facilities Division. Y.Y. was supported by the BES DOE, through the EPSCoR, Grant No. DE-FG02-08ER46528 as well as the Scientific User Facilities Division.

\section{Appendix: DFT calculation}

For UC, the electronic structure within the density functional theory (DFT) was obtained using the Quantum ESPRESSO package~\cite{09_giannozzi}. The calculation was performed using a plane-wave basis set and an ultrasoft pseudopotential~\cite{90_vanderbilt} optimized in a RRKJ scheme~\cite{90_rappe}. The uranium pseudopotential was obtained from an ionized electronic configuration: ${6p^6}{6d^1}{5f^3}{7s^1}$ with cutoff radii equal to 3.5 atomic units (a.u.), 1.7 a.u., 2.6 a.u. and 1.6 a.u. for $s, p, d$ and $f$ angular momentum. The electronic levels deviate from the all-electron ones by less than 0.1 meV. We used the Perdew, Burke, Ernzerhof~\cite{96_perdew} exchange-correlation functional. The Brillouin zone (BZ) summations were carried out over a $4\times4\times4$ supercell. The electronic smearing with a width of 0.02 Ry was applied according to the Methfessel-Paxton method. The plane wave energy and charge density cut-offs were 73 Ry and 1054 Ry respectively, corresponding to a calculation accuracy of 0.2 mRy/atom. The carbon atom potential was obtained from the total energy modification of a $2\times2\times2$ supercell when one carbon atom was shifted from the equilibrium position in the [100], [110], or [111] directions and the remaining atoms were held fixed in their equilibrium positions.

\end{document}